\documentclass[11pt]{article}
\usepackage[cp1251]{inputenc}

\usepackage{amsmath,amssymb,amsthm}

\usepackage{multicol}

\usepackage{color}
\usepackage{graphicx}
\renewcommand{\dfrac}[2]{{\displaystyle \frac{#1}{#2}}}
\allowdisplaybreaks

\date{}

\title{\textbf{Generalized quantum mechanical two Coulomb centers problem
(Demkov problem)}}

\author{A.~M.~Puchkov $^{\,\flat}${\footnote{putchkov@mail.ru, putchkov@yahoo.com}}, A.~V.~Kozedub
$^{\,\sharp}$ {\footnote{alexey.kozhedub@mail.ru}},
E.~O.~Bodnia$^{\,\flat}${\footnote{evgeniya.bodnya@cern.ch}}
\bigskip\\
$^{\,\flat}$ Theoretical Department, Institute of Physics, St.
Petersburg State University,\\
198904, Petergof, St. Petersburg, Russia\\
$^{\,\sharp}$ Department of Computational Physics, Faculty of
Physics, St. Petersburg State University, \\
 198904, Petergof, St.
Petersburg, Russia}

\begin{document}

\maketitle

\begin{abstract}

We present a new exactly solvable quantum problem for which the
Schr\"{o}dinger equation allows for separation of variables in
oblate spheroidal coordinates. Namely, this is the quantum
mechanical two Coulomb centers problem for the case of imaginary
intercenter parameter and complex conjugate charges is considered.
Since the potential is defined by the two-sheeted mapping whose
singularities are concentrated on a circle rather than at separate
points, there arise additional possibilities in choice of boundary
conditions. Detailed classification of the various types of
boundary-value problems is given. The quasi-radial equation leads
to a new type of boundary value problems which was never
considered before. Results of the numerical calculations allowing
to draw conclusions about the structure of the energy spectrum are
shown. Possible physical applications are discussed.

\end{abstract}

PACS number(s): {\bf{03.65.Ge\,, 12.39.Pn\,, 31.15.-p\,, 31.90 +
s}}


\bigskip
\noindent
{\Large Dedicated to the memory of Professor Dr. Yu.~N.~Demkov\\
(12. 04. 1926 -- 15. 11. 2010)}

\newpage

\section{Introduction}

The quantum mechanical two Coulomb centers problem $(Z_{1}eZ_{2})$
plays a fundamental role in various questions of atomic physics,
quantum chemistry of diatomic molecules and collision theory. The
problem is to determine wave functions and terms of an electron
moving in the field of two fixed charges $Z_{1}$ and $Z_{2}$
located at the distance $R$ from each other. The Hamiltonian of
the system, in atomic units $(\hbar = m_{e} = e = 1),$ is given by
\begin{equation}
\label{e1}
{\widehat H} = -\dfrac{1}{2}\Delta  -
\dfrac{Z_{1}}
{\left|
\bf{r} + \frac{\bf{R}}{2}
\right|} -
\dfrac{Z_{2}}
{\left|
\bf{r} - \frac{\bf{R}}{2}
\right|} \,.
\end{equation}
Due to high symmetry of (\ref{e1}), the corresponding
Schr\"{o}dinger equation allows for the separation of
variables in prolate spheroidal coordinates. As early as in 1930s,
Jaffe \cite{jaffe}, Baber and Hasse \cite{baber-hasse} offered
expressions for one-dimensional eigenfunctions in the form of
series whose coefficients satisfy three-term recurrence relations
\cite{kom_pon_sla}. Thereafter, based on these expansions,
algorithms were developed to calculate the terms with relative
accuracy of $10^{-12},$ and the wave functions $\sim 10^{-10}$
\cite{kom_pon_sla}, \cite{truskova}. All of this led to considering
various generalizations of the problem $(Z_{1}eZ_{2})$ (see, for
example, \cite{helf_hart1}).

Application of the two Coulomb centers problem in the atomic and
molecular scattering theory is based on the fact that the motion
of the electrons and nuclei can be considered adiabatically due to
large difference in their masses. In other words, the potential
curves $E(R)$ or quasimolecular terms are introduced for the
colliding particles. These curves are analytic functions of the
internuclear distance $R.$ Transitions between two terms
$E_{1}(R)$ and $E_{2}(R)$ are related to their common
complex branch point $R_{c}$ in the vicinity of which the energy
surface looks like a corkscrew:
$$
\Delta E(R) = E_{2}(R) - E_{1}(R) \sim \text{const} ( R -
R_{c})^{1/2} \,.
$$
In case of slow collisions $(v \ll 1 a.e.),$  the branch points
located near the real axis play a dominant role. Then, in the
framework of the adiabatic approximation, it is possible to obtain
a simple expression for the transition probability in such
inelastic processes as ionization \cite{dem_kom}. More recently,
in the 1980s, in connection with needs of physics of thermonuclear
fusion, a range of velocities $v \approx 1 a.e.$  attracted
interest, that is, there appeared a need to study positions of the
terms singular points in the two Coulomb centers problem,
throughout the complex plane of the intercenter (internuclear)
distance $R.$ For the first time, such calculations were started
by E.~A.~Solov'ev \cite{sol1} and then continued with his
coauthors \cite{sol2}, \cite{sol3}. The main result of these
studies is that the various types of "hidden" quasicrossings of
the terms were found, and approximate expressions relating
quasicrossing parameters to quasi-molecule characteristics and
quantum numbers were obtained. Note that in the works
\cite{helf_hart1} and \cite{sol1}--\cite{sol3}, the consideration
was conducted in prolate spheroidal coordinates. In other words, a
solution of the problem $(Z_{1}eZ_{2})$ for real $R$ was taken as
a basis for the generalizations, and the terms for complex $R$
were obtained by their analytic continuation from the real axis.

In this work, we consider the generalization of a different kind.
Some time ago, Yu.~N.~Demkov expressed the following idea [private
communication]. In going to an imaginary parameter $R$ and
complex-conjugate charges in the Hamiltonian (\ref{e1}), it
remains Hermitian. The Schr\"{o}dinger equation in this case will
also allow a separation of variables in oblate spheroidal
coordinates but not in prolate. Because the potential is
two-sheeted, there are additional possibilities in choice of
boundary conditions and formulation of boundary-value problems.
Thus, a new exactly solvable quantum problem (integrable system),
whose solution can not be reduced to the special case
\cite{sol1}--\cite{sol3}.

The paper is organized as follows. In the next section, we briefly
describe the properties of the potential relating to our problem.
The third part deals with the separation of variables and the
formulation of boundary-value problems. Particular attention
should be paid to appearance of a new type of the boundary-value
problems for the quasi-radial equation. Such type of the problems
has never been considered previously, which is connected to a
configuration of the equation singular points and configuration of
the region in which the eigenfunctions are found. Description of
these problems solution is the subject of a separate mathematical
work, so in this article we restrict ourselves to presenting the
results of numerical calculations. In the fourth part, the
asymptotics of the eigenfunctions and terms are presented. In
section 5, the results of the numerical calculations are given.
In Conclusions we discuss possible physical applications.

Note also that the classical analogue of our problem is known as
{\it {the generalized two fixed centers problem}} and refers to
celestial mechanics \cite{demin}. For the first time, the problem
arose in 1961, when it was necessary to take into consideration
the effect of non-sphericity of the Earth's gravitational field on
satellites trajectories.

\section{The potential and its properties}

The substitution of $R \to \imath R,\;\;Z_{1} \to q_{1} +
\imath q_{2},\;\;Z_{2} \to q_{1} - \imath q_{2}$ for the Hamiltonian
(\ref{e1}), where $q_{1},\, q_{2},\, R $ are real numbers, reduces
it to a Hermitian operator again, inasmuch as the last two terms
are complex-conjugate. This operator can be considered as a new
Hamiltonian whose coordinate part we will call {\it{the potential
of the generalized two Coulomb centers problem.}} For more
clearness, we use the Cartesian coordinate system. Then
\begin{equation}
\label{a1}
V =
\dfrac{q_1+\imath q_2}{\sqrt{x^2+y^2+\left(z-\imath \dfrac{R}{2}\right)^2}}+
\dfrac{q_1-\imath q_2}{\sqrt{x^2+y^2+\left(z+\imath \dfrac{R}{2}\right)^2}}\,,
\end{equation}
It is evident that the parameters $q_1$ and $q_2$ appear in the
expression (\ref{a1}) linearly, so we represent it as a sum
\begin{equation}
\label{a2}
V =  V_1(x,y,z;q_1,R) + V_2(x,y,z;q_2,R).
\end{equation}
Each member of the sum (\ref{a2}) has the following types of
symmetry:
\begin{enumerate}
  \item[1)]
  symmetry with respect to rotations around the $z-$axis by an arbitrary angle,

  \item[2)]
   symmetry (antisymmetry) with respect to reflection in the $xy$ plane:
   $$ (x,y,z) \mapsto  (x,y,-z), $$

  \item[3)]
  symmetry (antisymmetry) with respect to inversion:
  $$ (x,y,z) \mapsto(-x,-y,-z), $$

  \item[4)]
  scale symmetry:
    $$ x, y, z, R, q_{1}, q_{2} \mapsto \lambda x, \lambda y, \lambda
      z, \lambda R, \lambda q_{1}, \lambda q_{2} \;\;\;\;\; \lambda \in
      \{\mathbb{R}\} \backslash {0}. $$
\end{enumerate}
The property 3) is a consequence of 1) and 2). The last property
indicates that there are only two non-trivial parameters in the
generalized quantum mechanical two Coulomb centers problem: $R$
and ratio of $q_{1}/q_{2}$ or $q_{2}/q_{1}.$

Now, let us turn our attention to the fact that the expression
(\ref{a1}) defines two-sheeted mapping, which is singular on the
circle $C: x^2+y^2=R^2/4,\;\;z=0,$ and not at the points
$z_{\mathrm{1,2}} = \pm R/2$ as it was in the problem
$(Z_{1}eZ_{2}).$ Thus, the space in which the wave functions will
be determined becomes two-sheeted, and the potential $V$ does not
already allow a simple electrostatic interpretation.

Way out of this difficulty is the following. {\it First,} the
regular branches (sheets) can be glued along the singular circle
$C$ to a certain analog of the Riemann surface and the potential
$V$ can be considered as the electrostatic potential on the
extended space. {\it Second,} if we use the fact that any multiply
connected space can be made simply connected by inserting proper
barriers, and keep one or another branch in (\ref{a1}) fixed, we can
interpret the potential $V,$ in such space with the barrier, as a
certain electrostatic potential. It should be emphasized that the
topological considerations allow considerable arbitrariness in
choosing form of the barrier, it is only important that its
boundary coincides with the singular circle $C.$ However, the
requirement of the variables separation (as shown below) leaves
only three basic variants. They are the circle $C_{1}: x^2+y^2
\leqslant R^2/4,\;\;z=0,$ its exterior  $C_{2}: x^2+y^2 \geqslant
R^2/4,\;\;z=0$ and their union $C_{1} \bigcup C_{2},$ that is, the
entire $xy-$plane.

If the branches (\ref{a1}) are placed symmetrically relative to
the top and bottom sides of the barrier: $V_{+} = V_{-},$ we get
an analog of the simple-layer potential; when the branches are
placed antisymmetrically: $V_{+} = - V_{-},$ we get an analog of
the double-layer potential.

Now, let us agree on the terminology: in future, the spectral
problem in the extended space will be referred to as
{\it{two-sheeted problem,}} and in the ordinary space with a
barrier - {\it{one-sheeted problem.}}

\section{Separation of variables and formulation of boundary-value problems}

It is known \cite{kom_pon_sla} that the potentials in which the
Schr\"{o}dinger equation is separable in oblate spheroidal
coordinates $(\xi, \eta,\varphi)$ must be presented in the form
of:
$$ V = - \dfrac{2}{R^{2}} \left\{
\dfrac{a(\xi)-b(\eta)}{\xi^2 + \eta^2} + \dfrac{c(\varphi)}{(\xi^2
+ 1)(1 - \eta^2)} \right\}. $$
The connection of these coordinates
with Cartesian coordinates is given by the following relations:
\begin{equation}
\label{a3} x=\frac{R}{2}\sqrt{(\xi^2+1)(1-\eta^2)}\cos{\varphi}\,,
\; y=\frac{R}{2}\sqrt{(\xi^2+1)(1-\eta^2)}\sin{\varphi}\,, \;
z=\frac{R}{2}\xi\eta\,,
\end{equation}
where the variables domains $D$ is traditionally chosen by one of the two
alternative methods:
\begin{equation}
\label{a4}
a)\;\;\;  \xi \in [0,\infty),\;\;\; \eta \in [-1,1],\;\;\; \varphi
\in [0,2\pi);
\end{equation}
\begin{equation}
\label{a5}
b)\;\;\; \xi \in (-\infty, \infty),\;\;\; \eta \in [0,1],\;\;\;
\varphi \in [0,2\pi).
\end{equation}
Note that the cases (\ref{a4}) and (\ref{a5}) are related ro the
one-sheeted problem, since the relation (\ref{a3}) determines a
biunivocal mapping (bijection) $f:D \to {\mathbb R}^{3}.$

If we proceed to consider the two-sheeted problem, the variables
domain $\tilde D$ should be chosen as follows:
\begin{equation}
\label{a6} c)\;\;\;  \xi \in (-\infty, \infty)\,,\;\;\; \eta \in
[-1,1]\,,\;\;\; \varphi \in [0,2\pi)\,.
\end{equation}
Then the mapping $f:\tilde D \to {\mathbb{R}}^{3}$ will be
single-valued, and the inverse $f^{-1}: {\mathbb{R}}^{3} \to
\tilde D$ -- double-valued.

Let us change the variables in (\ref{a1}) according to (\ref{a3})
and recall that the potential of the generalized two Coulomb
centers problem can be constructed in different ways. A simple
analysis shows that there exists a total of nine variants of such
structures allowing the separation of variables, at that the most
physically meaningful are three of them:
\begin{equation}
\label{a7} V = - \dfrac{4(q_{1}\xi + q_{2}\eta)}{R(\xi^{2} +
\eta^{2})}\,,
\end{equation}
\begin{equation}
\label{a8} V = - \dfrac{4(q_{1}\xi + q_{2}\eta \;
\mbox{sign}(\eta))}{R(\xi^{2} + \eta^{2})}\,,
\end{equation}
\begin{equation}
\label{a9} V = - \dfrac{4(q_{1}\xi \; \mbox{sign}(\xi) +
q_{2}\eta)}{R(\xi^{2} + \eta^{2})}\,,
\end{equation}

The variant (\ref{a7}) refers to the double-sheeted problem with
the variables domain $\tilde D,$ as well as to one-sheeted problem
with the domain $D$ according to (\ref{a4}) or (\ref{a5}). When
the domain $D$ is determined by (\ref{a4}), the impermeable
barrier must be the circle $C_{1}.$ Then $V_{1}$ is interpreted as
analog of the simple-layer potential, and $V_{2}$ -- double-layer.
If the domain $D$ is determined by (\ref{a5}), the barrier must be
the exterior of the circle $C_{2}.$ Then $V_{1}$ is interpreted as
analog of the double-layer potential, and   $V_{2}$ --
simple-layer. The variants (\ref{a8}) and (\ref{a9}) are
derivatives from (\ref{a7}) and refer to one-sheeted problems
where $V_{1}$ and $V_{2}$ are analogs of the simple-layer
potential. In the case of (\ref{a8}), the domain $D$ is determined
by (\ref{a4}), and in the case of (\ref{a9}), it is determined by
(\ref{a5}).

It should be noted that if we consider the generalized quantum
mechanical two Coulomb centers problem in the half-space, the
barrier must be $xy-$plane. The domain $D$ is determined as
follows:
$$
d)\;\;\;  \xi \in [0,\infty),\;\;\; \eta \in [0,1],\;\;\; \varphi
\in [0,2\pi)\,.
$$
In this case, differences between (\ref{a7}), (\ref{a8}) and
(\ref{a9}) vanish. Let us follow the variable separation procedure
and the formulation of boundary-value problems by the example of
(\ref{a7}), since this variant is basic.

Let us represent the wave function $\Psi_j,$  corresponding to the
term $E_{j}(R),$ in the form of
\begin{equation}
\label{a10} \Psi_{j} = \Psi_{kqm}(\xi, \eta, \varphi; R) =
N_{kqm}(R) X_{mk}(\xi; R) Y_{mq}(\eta; R) e^{i m \varphi} \,,
\end{equation}
where the multiindex $j = \lbrace kqm \rbrace$ denotes the quantum
number set in which $k$ and $q$ coincide with the numbers of zeros
of the corresponding functions in the variables $\xi$ and $\eta,$
and the number $m$ takes the values $0,\pm 1, \pm 2, \dots .$
The normalization constant $N_{kqm}(R)$ is determined by the
condition
$$ \int \limits_{V}{\Psi_{kqm}^{\ast}(\xi, \eta, \varphi;
R) \Psi_{k'q'm'}(\xi, \eta, \varphi; R) d V} = \delta_{k
k'}\delta_{q q'}\delta_{m m'}\,,
$$
where  $dV = \frac{R^3}{8}(\xi^2 + \eta^2)d\xi d\eta d\varphi$ --
is a volume element in the oblate spheroidal coordinates. After
substituting (\ref{a7}) and (\ref{a10}) ) in the Schr\"{o}dinger
equation
$$
\Delta \Psi + 2\left(E - V \right)\Psi = 0
$$
we obtain the ordinary differential system
\begin{equation}
\label{a11}
\dfrac{d}{d\xi}(\xi^{2}+1)\dfrac{d}{d\xi}X_{mk}(\xi;R) -
\left[\lambda_{mk}^{(\xi)}+p^{2}(\xi^{2}+1) - a\xi - \dfrac{m^{2}}{\xi^{2}+1}
\right]X_{mk}(\xi;R) = 0 ,
\end{equation}
\begin{equation}
\label{a12}
\dfrac{d}{d\eta}(1-\eta^{2})\dfrac{d}{d\eta}Y_{mq}(\eta;R) +
\left[\lambda_{mq}^{(\eta)}+p^{2}(1-\eta^{2})+b\eta -
\dfrac{m^{2}}{1-\eta^{2}} \right]Y_{mq}(\eta;R) = 0 .
\end{equation}
Here  $p^{2}_{j} = -\dfrac{E_{j}R^{2}}{2}\; \; \;(p>0),\; \; \; a
= 2q_{1}R,\;\;\; b = - 2q_{2}R,$ at that $p$ has the meaning of
the energy parameter; $a$ and $b$ are the charge parameters;
$\lambda_{mk}^{(\xi)} = \lambda_{mk}^{(\xi)}(p,a)$ and
$\lambda_{mq}^{(\eta)} = \lambda_{mq}^{(\eta)}(p,b)$ are the
separation constants.

The equations (\ref{a11}) and (\ref{a12}) supplemented by the
boundary conditions form the boundary value problems that must be
solved simultaneously, and the energy spectrum can be obtained
from the condition
\begin{equation}
\label{a13}
\lambda_{mk}^{(\xi)}(p,a) = \lambda_{mq}^{(\eta)}(p,b).
\end{equation}
The general theory of Sturm-Liouville-type one-dimensional
boundary problems implies that the quantum numbers $k,\;q,\; m$
remain constant for the continuous variation of the intercenter
parameter $R,$ and the eigenvalues $\lambda_{mk}^{(\xi)}(p,a)$ or
$\lambda_{mq}^{(\eta)}(p,b)$ are non-degeneracy. Consequently, if
the solution of equation (\ref{a13}) exists, it is unique.

Let us now discuss the formulation of boundary-value problems. On
the one hand, it follows from the most common requirements for the
wave function that
\begin{equation}
\label{a14} \Psi_{j} \in {\cal L}_{2}
\left({\mathbb{R}}^{3}\right) \subset {\cal L}_{2}^{(\xi)}\left(
\mathbb{R} \right) \cup {\cal L}_{2}^{(\eta)}\left( [-1,1] \right)
\cup {\cal L}_{2}^{(\varphi)}\left( [0,2\pi) \right)\,.
\end{equation}
On the other hand, in the spatial domain where the potential
becomes infinite, the particle can not penetrate at all, that is,
there must be $\Psi_{j} = 0.$ The continuity of $\Psi_{j}$
requires that $\Psi_{j}$ becomes zero on the boundary of this
domain; generally speaking, in this case the derivative of
$\Psi_{j}$ has a jump \cite{landau}. Thus, in the one-sheeted
problem, where the circle $C_{1}$ is the barrier for the wave
function, we have:
\begin{equation}
\label{a15}
\Psi_+|_{C_{1}}=\Psi_-|_{C_{1}}=0\,.
\end{equation}
The normal derivative must jump
\begin{equation}
\label{a16} \left. \frac {\partial \Psi}{\partial n_+}
\right|_{C_1} - \left. \frac {\partial \Psi}{\partial n_-}
\right|_{C_1} = g(\eta;q_1,q_2,R)\,,
\end{equation}
where $g(\eta;q_1,q_2,R)$ is a smooth function of  $\eta$ and
parameters $q_1$, $q_2$ and $R.$ From the conditions (\ref{a14})
and (\ref{a15}), we obtain the boundary conditions for
quasi-radial function
\begin{equation}
\label{a17}
X_{m k}(0;R)=0, \quad | X_{m k}(\xi;R)  | \xrightarrow[\xi \to
\infty]{} 0.
\end{equation}
In the quasi-angular equation (\ref{a12}) the boundary points are
simultaneously singular, so to satisfy the condition (\ref{a14}),
it should be required that the function be limited in them:
\begin{equation}
\label{a18}
| Y_{m q}(\pm 1;R)|<\infty.
\end{equation}
If (\ref{a17}) and (\ref{a18}) are satisfied, then (\ref{a15})
also holds, and the condition (\ref{a16}) is satisfied
automatically.

In the one-sheeted problem with the barrier $C_{2},$ it is
necessary to impose on the wave function the following condition:
\begin{equation}
\label{a19}
\Psi_{+}|_{C_2}=\Psi_-|_{C_2}=0\,.
\end{equation}
For the normal derivative we obtain
$$
\left. \frac {\partial \Psi}{\partial n_+} \right|_{C_2} - \left.
\frac {\partial \Psi}{\partial n_-} \right|_{C_2} =
h(\eta;q_1,q_2,R)\,,
$$
where $h(\eta;q_1,q_2,R)$ is a smooth function of  $\eta$ and
parameters $q_1$, $q_2$ and $R.$ From the (\ref{a19}), we obtain
the boundary conditions for the quasi-radial and quasi-angular
functions:
\begin{equation}
\label{a20}
| X_{m k}(\xi;R)  | \xrightarrow[\xi \to \pm \infty]{} 0, \quad
Y_{m q}(0;R)=0, \quad | Y_{m q}(+1;R)|<\infty\,.
\end{equation}
In the two-sheeted problem no barriers appear, but there is
a circle $C$ on which the potential is singular, therefore, the
following condition must be satisfied:
\begin{equation}
\label{a21}
\Psi|_C=\Psi(0,0,\varphi;R)=0\,,
\end{equation}
which is in contradiction with the expression (\ref{a10}). Indeed,
if the expressions (\ref{a10}) and (\ref{a21}) are valid, we have
the alternative:
\begin {enumerate}
\item $X_{m k}(0;R)=0\,,$ or $Y_{m q}(0;R)=0;$

\item $X_{m k}(0;R)=0$ and $Y_{m q}(0;R)=0$ simultaneously,
\end{enumerate}
which, generally speaking, does not follow from any physical
considerations. The contradiction arises particularly acute when
we consider the ground state.

It is apparent that both cases are not satisfied. The way out
of the situation is to separate factors (factorization)
\begin{equation}
\label{a22}
\Psi_j= \overline N_{k q m}(R)({\xi}^2+{\eta}^2)
\overline X_{mk}(\xi;R)
\overline Y_{m q}(\eta;R)
e^{i m \varphi}\,.
\end{equation}
Here, the overline means that we are dealing with one-dimensional
functions of the two-sheeted problem. Although the functions
satisfy (\ref{a11}) and (\ref{a12}), but they essentially
different from the corresponding functions of the one-sheeted
problem.

The boundary conditions in the two-sheeted problem are formulated
as follows:
\begin{equation}
\label{a23}
|\overline X_{m k}({\xi};R)| \xrightarrow[\xi \to \pm \infty]{} 0,
\quad | \overline Y_{m q}(\pm 1;R)|<\infty,
\end{equation}
because now the boundary points are singular.

Let us now discuss the specific of the boundary-value problems
related to the quasi-radial equation. In the problem
$(Z_{1}eZ_{2})$ the corresponding equation was considered on the
interval $[1, \infty),$ that is, between the singular points.
Then, using the Jaffe transformation \cite{jaffe} for the
independent variable $x = (\xi - 1)/(\xi + 1),$ the interval is
transferred to the unit segment, and the additional singular point
-1 goes to $-\infty$ and does not effect the convergence of
series representing the eigenfunction. In our problem, the
equation (\ref{a11}) on the complex $\xi-$plane has three singular
points: two finite regular point $\xi_{1} = +\imath, \xi_{2} =
-\imath$ and one irregular at infinity.

It is clear that the points $\xi_{1}$ and $\xi_{2}$ are "on different sides"
of the real axis, and so any attempt to present
$X_{mk}(\xi;R)$ in the form of series faces the problem of the
circle of convergence. Use of standard techniques, such as the
transformation of Jaffe \cite{jaffe} or Jaffe-Lay \cite{slav_lay},
are no longer possible here, and therefore it is necessary to look
for a more general method including the standard methods as
special cases. Note that the problem of the circle of convergence
will always appear in the boundary-value problems after separation
of variables in the Schr\"{o}dinger equation in oblate spheroidal
coordinates (it is typical for these coordinates).

In conclusion of this paragraph we note that the separation of the
factors in (\ref{a22}), without changing the structure of the
equations (\ref{a11}) and (\ref{a12}), can affect the convergence
of the series for the one-dimensional functions.

\section{Asymptotic behavior of the eigenfunctions and terms as $R \to 0.$}

In the context of possible practical applications of the model
under consideration, the pattern of the terms is of the most
interest. In order to correctly identify the individual terms it
is necessary to know the initial approximation or asymptotic of
$E_{j}(R)$ as $R \to 0.$ In this limit, the difference between the
one-sheeted and two-sheeted problems disappears and, in addition,
the spherical symmetry arises, therefore it is natural to go to
spherical coordinates $(r,\vartheta,\varphi).$

Let us expand the potential (\ref{a1}) in powers of the small
parameter $(R/r)$ using the formula (see, for example,
\cite{grR}):
$$
\dfrac{1}{\sqrt{1 - 2tu + t^2}} =
\sum_{k=0}^{\infty}{t^{k}P_{k}(u)}\,, \;\;\;
|t| < min|u \pm \sqrt{u^2 - 1}|\,,
$$
where $P_{k}(u)$ is Legendre polynomials. As a result of the
transformations we obtain the expression
\begin{equation}
\label{c1}
V =
\dfrac{2q_{1}}{r}\sum_{n=0}^{\infty}
(-1)^{n}
\left(
\dfrac{R}{2r}
\right)^{2n}
P_{2n}(\cos{ \vartheta}) +
\dfrac{2q_{2}}{r}\sum_{n=0}^{\infty} (-1)^{n+1} \left(
\dfrac{R}{2r} \right)^{2n+1} P_{2n+1}(\cos{ \vartheta}) \,,
\end{equation}
where odd multipole moments associated with the parameter $q_{1},$
and the even multipoles with $q_{2}.$ It is clear that only one
term $2q_{1}/r$ in (\ref{c1}) does not vanish as $R \to 0.$ This
term is naturally chosen as the unperturbed potential while the
perturbation will be the first term of the second sum $-
q_{2}R/r^{2} cos \vartheta .$ Then the terms $E_{j}(R)$ must
continuously transit to $N^2-$fold degenerate energy levels of the
hydrogen atom with a charge $2q_{1}:$
\begin{equation}
\label{c2}
E_{j}(R) \xrightarrow [ R \to 0]{} E_{Nlm} = -
\dfrac{2q_{1}^{2}}{N^{2}}\,,
\end{equation}
where $i = \lbrace N l m \rbrace-$is a set of spherical quantum
numbers which are related to the numbers of zeros of the
one-dimensional eigenfunctions by the relations: $$ N = k + q + m
+ 1, \quad l = q+m\,.$$ The radial $X_{mk}(\xi;R)$ and angular
$Y_{mq}(\eta;R)$ parts of the eigenfunction $\Psi_{j}$ are reduced
to into the radial $R_{Nl}(r)$ and angular $Y_{lm}(\vartheta,
\varphi)$ parts of one-center problem.
\begin{equation}
\label{c3}
\Psi_{kqm}(\xi, \eta, \varphi; 0) =
N_{kqm}(0)X_{mk}(\xi;0)Y_{mq}(\eta;0)e^{i m \varphi} =
\tilde{N} R_{Nl}(r)Y_{lm}(\vartheta, \varphi)\,,
\end{equation}
that will be the correct function of the zero approximation.
Arguments similar to those in \cite{kom_pon_sla}, give the
following expression for the energy in the first non-vanishing
order:
\begin{equation}
\label{c4} E_{Nlm}(q_{1}, q_{2}, R) = - \dfrac{2q_{1}^{2}}{N^{2}}
- \dfrac{8q_{1}^2(q_{1}^{2}+q_{2}^{2})[l(l+1) - 3m^2]R^2}
{N^{3}l(l+1)(2l-1)(2l+1)(2l+3)} + O((R)^2)\,.
\end{equation}
Note that formula (\ref{c4}) reduces to Baber and Hasse's result
\cite{baber-hasse} by means of the substitution $q_{1} \to (Z_{1}
+ Z_{2})/2, \;\;\;  q_{2} \to \imath (Z_{1} - Z_{2})/2.$ The
calculation of the energy in the following order of the
perturbation theory becomes quite cumbersome and, in addition,
there one can expect logarithmic corrections similar to those obtained
in \cite{levina} for the problem $(Z_{1}eZ_{2}).$

\section{The results of numerical calculations}

The boundary-value problems, which are equations (\ref{a11}) and
(\ref{a12}) supplemented by the boundary conditions (\ref{a17}),
(\ref{a18}) and (\ref{a20}), have been solved numerically. As
noted above, the condition (\ref{a13}) for the fixed parameters
$a, b,$ and $j = \lbrace kqm \rbrace$ has a unique solution
\begin{equation}
\label{f1} p^{\ast}_{j}(2q_{1}R, - 2q_{2}R) =
\dfrac{R}{2}\sqrt{-2E_{j}}\,.
\end{equation}
Solving it for $E_{j},$ we find discrete spectrum. For the
classification of the terms, in addition to $j = \lbrace kqm
\rbrace,$ we will also use the set of spherical quantum numbers $i
= \lbrace N l m \rbrace.$ Let us keep the traditional
spectroscopic notation, when the numbers
$$l = 0, \; 1, \; 2, \;  3, \; 4, \; 5, \dots \quad \mbox{and} \quad
m = 0, \; 1, \; 2, \; 3,\dots$$ correspond to the letter rows $$L
= s,\; p, \; d, \; f, \; g, \; h, \dots\quad \mbox{and} \quad M =
\sigma, \; \pi, \; \delta, \; \phi, \dots$$ Therefore, in order to
make any conclusions about the structure of the energy spectrum in
the generalized quantum mechanical two Coulomb centers problem, at
least on a qualitative level, it is necessary to find a
meaningful collection of the special cases of $E_{j}(R)$  which
covers the transition and asymptotic (as $R \to 0\,, \; R \to
\infty$) domains.

Calculations with different combinations of $q_{1} = 1\,, \ldots
10$ and $q_{2} = 1\,, \ldots 10,$ show that such collection is
formed from the first ten $E_{j}(R)$ at $R \in [0; 20].$ It can
be explained by the fact that there is a scaling (see property 4)
of the potential (\ref{a1})) in our problem. Thus, in spite of the
fact that the total of about two thousand curves have been calculated,
the structure of the spectrum was determined in a very wide domain
of variation of $q_{1}, q_{2}$ and $R.$

\begin{figure}[H]
  \noindent\centering{
    \includegraphics[scale=0.85]{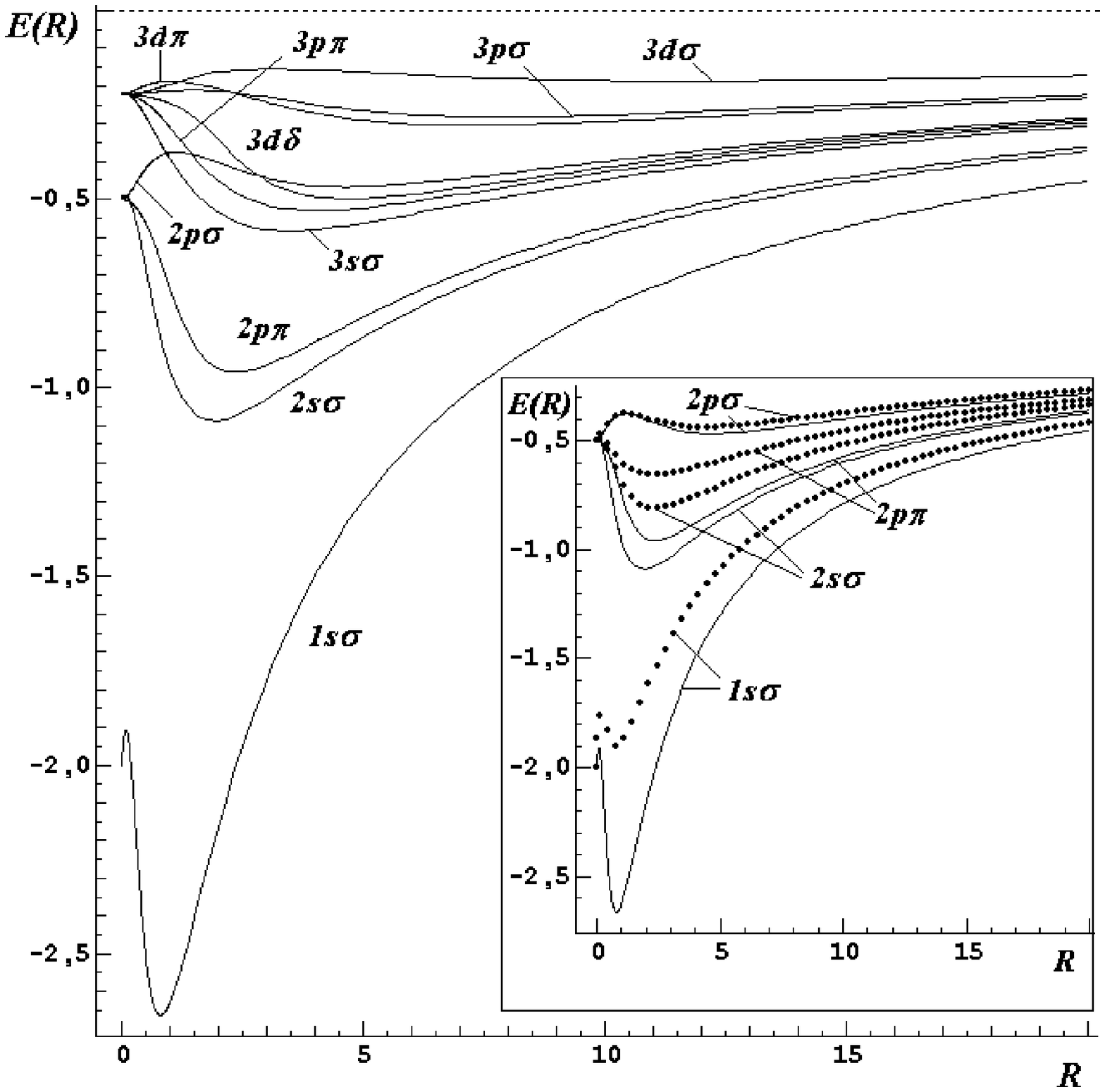}
  }
  \caption{Terms of system $q_{1}=1,\; q_{2}=3.$}
  \label{fig:1}
\end{figure}

\begin{figure}[H]
  \noindent\centering{
    \includegraphics[scale=0.65]{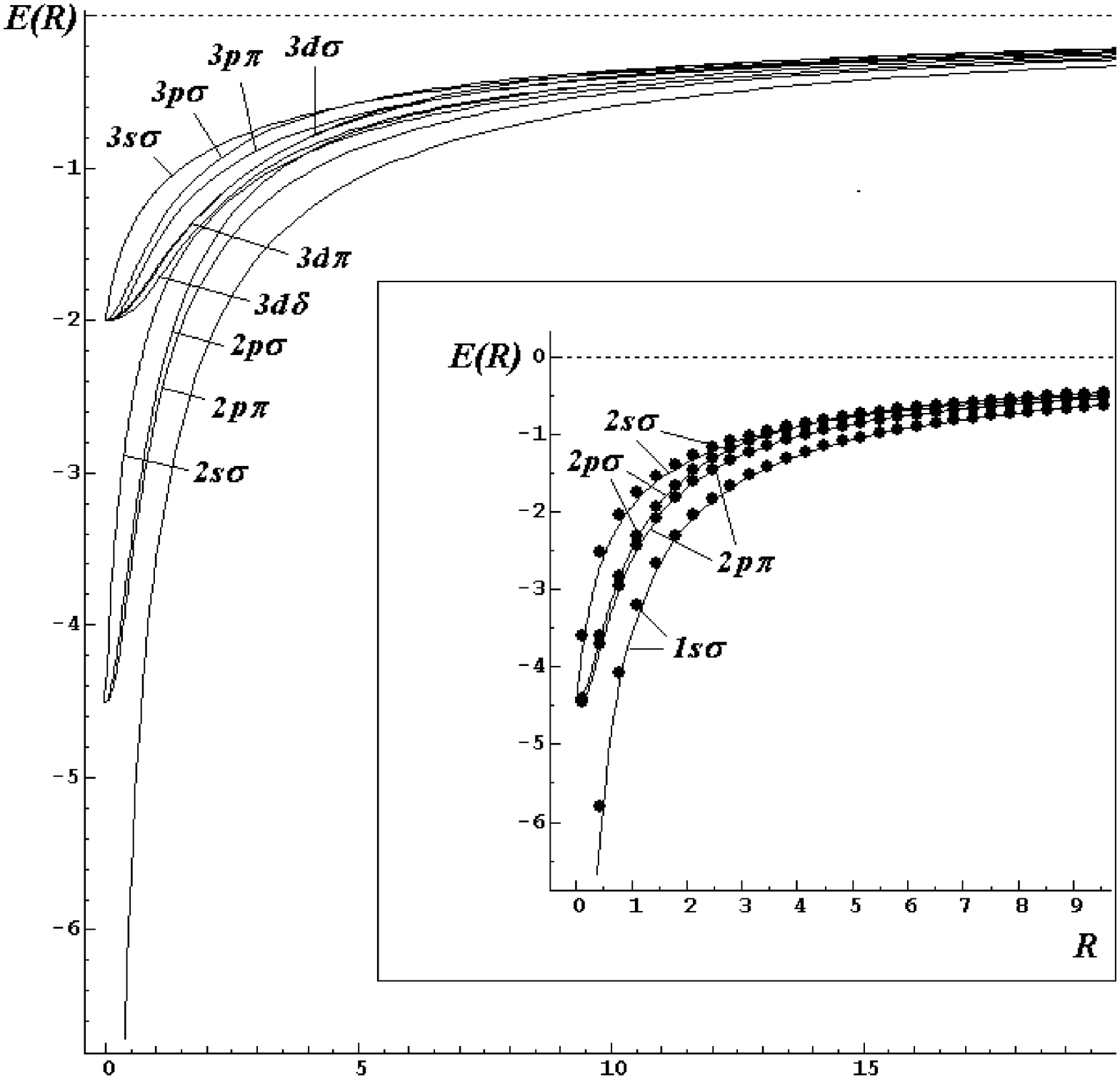}
  }
  \caption{Terms of system $q_{1}=3,\; q_{2}=1.$}
  \label{fig:2}
\end{figure}

Analysis of the curves shows that the pattern of the terms varies
considerably for $q_{1} < q_{2}$ and $q_{1} > q_{2}.$ This
statement becomes especially obvious when selecting two systems
differing from each other by the permutation $q_{1}
\leftrightarrow q_{2},$ for example, $q_{1}=1,  q_{2}=3$ (see Fig.
\ref{fig:1}) and $q_{1}=3, q_{2}=1$ (see Fig. \ref{fig:2}).

\begin{figure}[H]
  \noindent\centering{
    \includegraphics[scale=0.65]{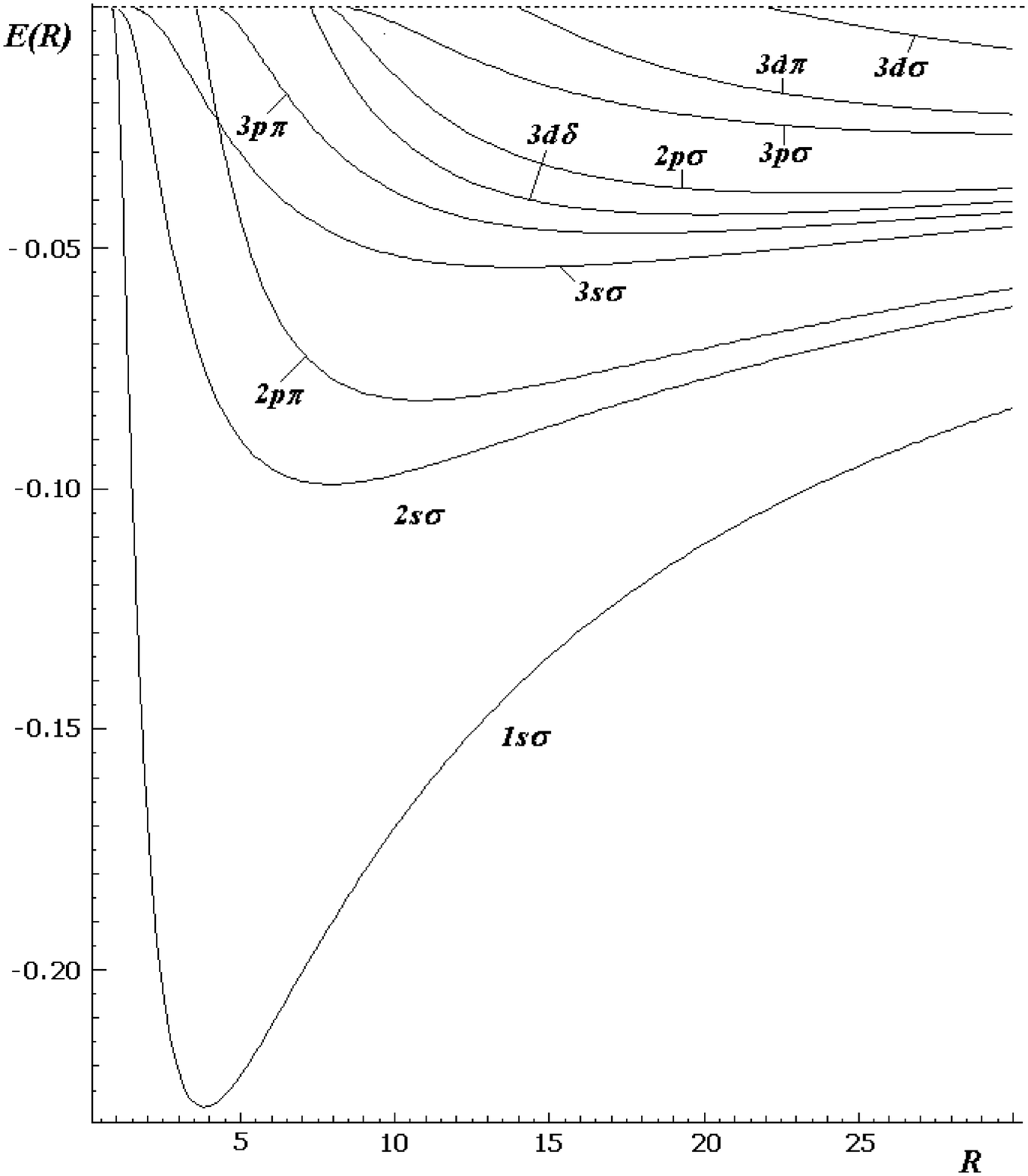}
  }
  \caption{Terms of system $q_{1}=0,\; q_{2}=1.$}
  \label{fig:3}
\end{figure}

First of all, pay attention to the fact that the terms in Fig.
\ref{fig:1} have distinct local minima at finite $R,$ whereas the
terms in Fig. \ref{fig:2} change rather aslope. The availability
of the minima of $E_{j}(R)$ at $q_{1} < q_{2}$ indicates the
stable states of a moving charged particle in such systems.

The case of $q_{1}=0$ requires a separate consideration. It is
well-known \cite{baz_zeldovich} that in the spherically symmetric
field $C r^{-2}$ with $C \leqslant - 1/4$ the energy levels
exponentially condense to the continuous spectrum border. One can
see a similar effect in the Fig. \ref{fig:3}. We stress that the
continuous spectrum starts at finite values of $R.$

Note that in all three figures the dashed lines depict
the border of continuous spectrum.

The question of the configuration interaction of the terms or
their quasicrossings requires special consideration. In the works
of E.~A.~Solov'ev \cite{sol1}--\cite{sol3}, the branch points were
found near the imaginary axis $R,$ so we can expect quasicrossings
in the spectrum of our problem.

\section{Conclusion}

We have considered a new exactly solvable quantum problem. In
distinction to the $(Z_{1}eZ_{2})$ problem, our problem allows for
various choices of boundary conditions. It gives us additional
opportunities in setting the boundary problems. We have shown that
the boundary problems associated with the quasi-radial equation
belong to a new type. The peculiarity of these problems is related
to the fact that the singular points of the differential equation
are on different sides of the region in which the eigenfunctions
are defined. Analytic representations of such eigenfunctions are
not known yet. Even in the famous paper \cite{rainwater} where an
analogous problem for spheroidal equations has arised, the
boundary problems were not duly considered and the relevant
discussions were largely avoided. Therefore, our paper can be
considered as an incentive for the mathematical physics experts to
analyse the problems of this type. In particular, it owuld be interesting
to apply the Nikiforov-Uvarov method
\cite{nikiforov-uvarov} and asymptotic iteration method (AIM)
\cite{ciftci1}, \cite{ciftci2} and \cite{boztosun-karakoc} to
these problems.

Note that in recent paper \cite{zhang-huangfu} an exactly solvable
problem was considered with a ring-shaped noncentral potential.
Singularities of this potential were located on a plane, instead of
a circle as in our work, and the separation of variables in the
Schr{\" o}dinger equation was performed in spherical coordinates.
Therefore, the Nikiforov-Uvarov method could be used for
determination of the eigenfunctions. Our boundary problems are
far more complex. Nikiforov-Uvarov method has never been applied
to such models.

We have shown for the first time that quantum problems can be
considered on Riemannian surfaces. Probably, this idea could give
some insights into the puzzle of confinement of quarks. At least,
one could consider models of \cite{matrasulov1},
\cite{matrasulov2} and \cite{majethiya} type on Riemannian
surfaces.

Numerical calculations have shown that for $q_{1} < q_{2}$  the
terms $E_{j}(R)$ do have minima at finite values of $R.$ It
implies that such systems possess stable states. We believe that
the Demkov problem may be viewed as a model for higher excited electron states
in the fields of various ring molecules (e.g., aromatic
hydrocarbons). In particular, the asymptotic formulas (\ref{c3})
and (\ref{c4}) may be of use for description of higher excited
states of benzene \cite{grubb} which have been observed
experimentally.
Exact quantum mechanical computation of these states is very
challenging from the technical side. Our work offers a possibility
of simple description not only for benzene but for a good number of ring molecules.

\end{document}